%% file: main.tex
\documentclass{article}

\input{macpack}

\usepackage{arxiv}
\usepackage{tabularx} 
\usepackage[]{inputenc} %
\usepackage{multirow}

\title{A Qualitative Analysis of Kernel Extension for Higher Order Proof Checking}

\author{
 Shuai Wang \\
 INRIA Rocquencourt, Paris, France\\
ILLC, University of Amsterdam, The Netherlands\\
  \texttt{shuai.wang.vu@gmail.com} \\
}

\begin{document}
\maketitle
\begin{abstract}
For the sake of reliability, the kernels of Interactive Theorem Provers (ITPs) are generally kept relatively small. On top of the kernel, additional symbols and inference rules are defined. This paper presents an analysis of how kernel extension reduces the size of proofs and impacts proof checking.\footnote{The author was supported by the MPRI-INRIA scholarship during this internship. The paper was presented in the student session of the European Summer School in Logic, Language, and Information (ESSLLI) in 2016.}
\end{abstract}


\section{Introduction}
\justify

Higher order logic is also known as simple type theory. It is an extension of simply typed $\lambda$-calculus with additional axioms and inference rules \cite{seven}. Interactive Theorem Provers (ITPs) of higher order logic have been playing an important role in formal mathematics, software verification and hardware verification. However, ITPs may have bugs and may lead to errors in proofs generated while not being apparent within the proof systems themselves. Also, proofs nowadays can be huge, making it difficult or even impossible to check by hand. For example, the Kepler Conjecture project took a team of scientists several years with many ITPs involved \cite{kepler}. The demand of reliability of such ITPs makes proof checking necessary, especially by proof checkers independent from the ITPs. Taking advantage of the similarity of the logic and design between some ITPs, OpenTheory \cite{opentheory} has developed a standard format for serialising proofs \cite{opentheory}. One way to verify these proofs (also known as proof articles) is to export them to the OpenTheory format followed by the proof checking process by Dedukti \cite{dedukti}.

The correctness of an ITP depends on its kernel where basic symbols and inference rules are defined \cite{harrison2006towards, myreen2013steps}. On top of the kernel, more symbols and corresponding inference rules are defined. The kernel of HOL Light takes equality as its only logical (term) symbol to keep the size of its kernel minimal. Some dependency analyses of the symbols of the HOL Light system show that, aside from equality, there is also much dependency on implication and universal quantification. In contrast, HOL4 takes conjunction, disjunction, implication, existential quantification and so on as primitive symbols. This paper presents HOLALA\footnote{The source code of HOLALA and the internship report can be found on Zenodo \cite{wang_2024_10624348}.}, an alternative version of HOL Light with a kernel extension of additional symbols and inference rules. More specifically, implication and universal quantification were taken primitive. This paper presents an experimental work on qualitative measurement of the impact of kernel extension with a concentration on proof checking efficiency.

This paper is organized as follows: Chapter \ref{ch:kernel} explains the kernel of HOL Light and Chapter \ref{ch:hacking} illustrates the design of HOLALA by extending the kernel of HOL Light. Following that is the update of Holide and Dedukti as well as proof checking and evaluation in Chapter \ref{ch:evaluation}.  


\section{HOL Light}
\label{ch:kernel}

Higher order logic is also known as simple type theory. It is a logic on top of simply typed $\lambda$-calculus with additional axioms and inference rules \cite{seven}. The type of a term is either an individual, a boolean type or a function type. A term is either a constant, a variable (e.g. $x$), an abstraction (e.g. $\lambda x.x$) or a well-typed application (e.g. $(\lambda x.x) y)$. The notation $x : \iota$ means that the term $x$ is of type $\iota$. Types are sometimes omitted for simplicity of representation.

\begin{table}[]
\centering
\begin{tabular}{ll}
type variables & $\alpha, \beta$ \\ 
type operators  & $p$   \\
types & $A, B::= \alpha \,|\, p(A_{1},\ldots,A_{n})$ \\
term variables & $x, y$\\
term constants  & $c$ \\
terms & $M, N ::= x \,|\, \lambda x: A. M \,|\, M N \,|\, c$ \\
\end{tabular}
\end{table}

HOL Light \cite{hollight} is an open source interactive theorem prover for higher order logic.  Its logic is an extension of Church's Simple Type Theory \cite{barendregt2013lambda} with polymorphic type \cite{hollight}. The kernel of HOL Light is an OCaml file where terms, types, symbols and inference rules are defined. Symbols and inference rules in the kernel are considered primitive. On top of the kernel, additional symbols are introduced and inference rules are derived. The kernel of HOL Light has only one primitive logical (term) symbol, the equality $(=)$\footnote{$s = t$ is a conventional concrete syntax for $((=) s t)$.}. The equality is of polymorphic type \cite{henkin1963theory} and plays three roles in HOL Light: definition, equivalence and bi-implication. 



\section{Kernel Extension}
\label{ch:hacking}

\begin{figure}[!ht]
 \centering
\begin{tikzpicture}[scale=0.6]

\tikzstyle{conjecture}=[rectangle]
\tikzstyle{title} = [rectangle]
\node[conjecture] (eq1) at (-3, 0) {=};
\node[conjecture] (t1) at (-4, 1) {$\top$};
\node[conjecture] (and1) at (-2, 2) {$\land$};
\node[conjecture] (imp1) at (-3, 3) {$\rightarrow$};

\node[conjecture] (forall1) at (-4, 3) {$\forall$};
\node[conjecture] (bot1) at (-5, 5) {$\bot$};
\node[conjecture] (or1) at (-4, 4) {$\lor$};

\node[conjecture] (exists1) at (-2, 4) {$\exists$};

\node[conjecture] (neg1) at (-3, 6) {$\neg$};

\draw[->] (t1) -- (eq1); 
\draw[->] (and1) -- (eq1);
\draw[->] (and1) -- (t1);
\draw[->] (imp1) -- (and1);
\draw[->] (imp1) -- (eq1);
\draw[->] (forall1) -- (eq1);
\draw[->] (forall1) -- (t1);
\draw[->] (exists1) -- (imp1);
\draw[->] (exists1) -- (forall1);
\draw[->] (or1) -- (imp1);
\draw[->] (or1) -- (forall1);
\draw[->] (bot1) -- (forall1);
\draw[->] (neg1) -- (bot1);
\draw[->] (neg1) -- (imp1);

\node[conjecture] (eq2) at (3, 0) {=};
\node[conjecture] (t2) at (5, 1) {$\top$};
\node[conjecture] (and2) at (2, 1.5) {$\land$};
\node[conjecture] (imp2) at (4, 0) {$\rightarrow$};
\node[conjecture] (forall2) at (2, 0) {$\forall$};
\node[conjecture] (bot2) at (1, 2) {$\bot$};
\node[conjecture] (or2) at (1, 1) {$\lor$};
\node[conjecture] (exists2) at (3, 1.5) {$\exists$};
\node[conjecture] (neg2) at (4, 3) {$\neg$};

\draw[->] (t2) -- (forall2); 
\draw[->] (t2) -- (imp2);
\draw[->] (or2) -- (forall2);
\draw[->] (or2) -- (imp2);
\draw[->] (and2) -- (imp2);
\draw[->] (and2) -- (forall2);
\draw[->] (exists2) -- (imp2);
\draw[->] (exists2) -- (forall2);
\draw[->] (bot2) -- (forall2);
\draw[->] (neg2) -- (imp2);
\draw[->] (neg2) -- (bot2);

\node[title] (title1) at (-3, -1) {HOL Light};
\node[title] (title2) at (3, -1) {HOLALA};

\end{tikzpicture}
\caption{Dependency Analysis}
\label{fg:logicdependency}
\end{figure}
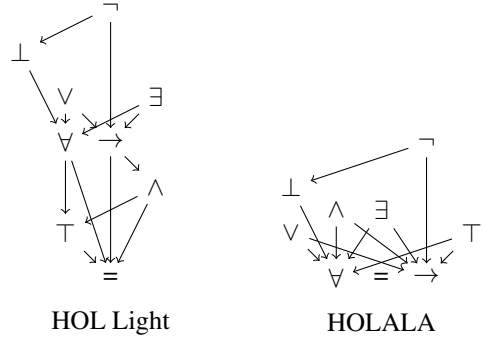

\begin{table*}[!ht]
\centering
\caption{Primitive and Axiomatic Definitions of Connectives and Constants Comparison}
\label{tb:logicdefine}
\begin{tabular}{lll}
          &  HOL Light & HOLALA    \\ \hline
\multicolumn{1}{l|}{=} & primitive            & primitive \\
\multicolumn{1}{l|}{$\rightarrow$}   &     $\lambda p q. p \land q \Leftrightarrow p$    & primitive         \\
\multicolumn{1}{l|}{$\forall$}          & $\lambda p. (p = \lambda x. \top)$  & primitive         \\
\multicolumn{1}{l|}{$\Leftrightarrow$}       & =                   & = \\
\multicolumn{1}{l|}{$\exists$}  & $\lambda p \forall q (\forall x. p x \rightarrow q) \rightarrow q$        &    $\lambda p \forall q (\forall x. p x \rightarrow q) \rightarrow q$       \\
\multicolumn{1}{l|}{$\top$}        & $\lambda p.p = \lambda p.p$                    & $\forall x. (x \rightarrow x)$         \\
\multicolumn{1}{l|}{$\bot$}         & $\forall p. p$                    & $\forall p. p$         \\
\multicolumn{1}{l|}{$\land$}    & $\lambda p q. (\lambda f. f p q) = (\lambda f. f \top \top)$      & $\lambda p q. (\forall x. (p \rightarrow ((q \rightarrow x) \rightarrow x)))$         \\
\multicolumn{1}{l|}{$\lor$}       & $\lambda p q. \forall r.(p \rightarrow r) \rightarrow ((q \rightarrow r) \rightarrow r)$  &   $\lambda p q. \forall r.(p \rightarrow r) \rightarrow ((q \rightarrow r) \rightarrow r)$   \\ 
\multicolumn{1}{l|}{$\neg$}   & $\lambda p. p \rightarrow \bot$ & $\lambda p. p \rightarrow \bot$ \\

\end{tabular}
\end{table*}

\begin{table*}[!ht]
\centering
\renewcommand{\arraystretch}{2}
\caption{Primitive Inference Rules of HOL Light \cite{hollight}}
\label{holrules1}
\setlength\extrarowheight{1pt}
\begin{tabular}{l|l}
 \hline
Structural                            &        \HOLASSUME        \\ \hline
\multirow{2}{*}{$\lambda$ Calulus}  &         \HOLABS          \\
                                    &          \HOLBETA        \\ \hline
\multirow{2}{*}{Instantiation}        &           \HOLINST        \\
                                    &          \HOLINSTTYPE        \\ \hline
\multirow{2}{*}{Bi-implication}     &         \HOLEQMP         \\
                                    &        \HOLDEDUCTAS         \\ \hline
\multirow{3}{*}{Equality}           &          \HOLREFL      \\
                                    &         \HOLMKCOMB         \\
                                    &         \HOLTRANS  \\\hline
\end{tabular}
\end{table*}

On top of the kernel, more logic connectives and constants are introduced. Figure \ref{fg:logicdependency} illustrates the dependency of these symbols based on their definition as in Table \ref{tb:logicdefine}. For example, the definition of $\exists$ depends on that of $\rightarrow$. Note that equality is in fact used when introducing every symbol but the graph omits such arrows for the sake of simplicity. It can be observed that logical connectives have much dependency on implication and universal quantification as well. This leads to the idea of introducing them as primitive symbols to reduce the depth of dependency and shorten  proofs without changing proof scripts. 

The kernel also includes ten primitive inference rules as in Table \ref{holrules1} (with types eliminated to keep the table small). On the base of the ten primitive inference rules, we introduce derived inference rules. The correctness of the proofs largely depends on the correctness of the kernel \cite{harrison2006towards, myreen2013steps}.

 The kernel of HOL-style ITPs are generally kept small for the sake of reliability. A kernel provides  primitive types, core inference rules and constants and various safe definitional mechanisms. If correctly implemented (assume the correctness of the meta-language),  the soundness of the ITP is guaranteed. Kernels vary from small ones (e.g. HOL Light and HOL Zero) to larger ones (e.g. HOL4). The more constants and inference rules taken primitive in the kernel, the harder it is to guarantee the soundness of the system. Although it is known to the HOL community that correctly expanding a kernel would lead to some efficiency gains, there is no qualitative measurement of this benefit. This paper shows how the extension of kernels would reduce the depth of dependency, leading to a reduction of the size of proofs and a speedup of proof checking without the loss of reliability. We introduce HOLALA, a modified version of (OpenTheory) HOL Light\footnote{(OpenTheory) HOL Light is HOL Light equipped with proof recording methods and exports proofs into proof packages, namely the article files. (OpenTheory) HOL Light also generates the standard library of the OpenTheory Repository. We refer to (OpenTheory) HOL Light as HOL Light in the rest of this paper for short (despite the differences in some detailed proofs in each systems and other aspects). } where the kernel consists of more logic symbols and their corresponding inference rules.  Different from HOL Light which takes equality as the only primitive symbol, HOLALA has an extended the HOL Light kernel with universal quantification and implication and their associated introduction and elimination rules ($\textit{MP}$, $\textit{GEN}$, $\textit{DISCH}$ and $\textit{SPEC}$). This was achieved by adding the universal quantifier and implication symbol to the kernel\footnote{Note that, similar to equality, universal quantification is also of polymorphic type.}. In addition, HOLALA also modified the definition of truth ($\top$), and conjunction ($\land$), making as many definitions of logic symbols as possible dependent on the universal quantifier and implication instead. The definition of symbols of HOLALA in comparison with HOL Light is shown in Table \ref{tb:logicdefine}. To summarise, Figure \ref{fg:logicdependency} shows a comparison of the dependency of symbols in HOL Light and HOLALA.  As a consequence, some derived inference rules were reproved. An immediate benefit of such changes is that the derived inference rules directly depending on inference rules of implication and universal quantification were shortened. For example, the conjunction introduction rule is expanded to 31 inference steps instead of 55 while recording. Similarly, the disjunction introduction rule takes 21 inference steps instead of 156. For this reason, proofs are expected to be shorter.


\begin{itemize}
  \item[] \HOLMPl
  \item[] \HOLGENl
  \item[] \HOLDISCHl
  \item[] \HOLSPECl
\end{itemize}






Although such changes lead to the reduction of proof size, users would lose the original definitions of the $\forall$ and $\rightarrow$. To fix proofs explicitly involving these two definitions, the definitions of the $\forall$ and $\rightarrow$ are proved as theorems after the introduction of the axiom of extensionality.

\section{Proof Checking and Evaluation}
\label{ch:evaluation}



\subsection{Extending Holide and Dedukti}
 In this project we employ Dedukti as the proof checker to verify the proofs generated by HOLALA. Cousineau and Dowek showed that Higher Order Logic can be embedded in the $\lambda\Pi$-calculus Modulo as well as other Pure Type Systems (PTS) \cite{gilles1}. This laid the foundation of Dedukti \cite{dedukti}, a universal proof checker. On top of Dedukti, Holide\cite{holide} was developed to transform proofs from a proof repository, namely the OpenTheory Repository \cite{opentheory}, to Dedukti. Following the extension of the logic kernel of HOL Light, there are some necessary changes to the existing translation of HOL Light's logic into Dedukti to accommodate this larger kernel. To deal with this update, the declaration of the universal quantifier and the implication together with their elimination and introduction inference rules were added to Holide as well as the input to Dedukti. The quantified terms would be translated as follows, with the notation of translation follows from the notation of \cite{holide}: 
 
 \begin{itemize}
  \item[]  $|\rightarrow |=imp$
  \item[]  $|(\forall_{A})|=forall |A|$
 \item[] $|\forall (M:A)| = forall |A| |M| $
 \item[]  $|M\rightarrow N|=imp|M||N|$, where


  \item[] imp: $term\,bool\rightarrow term\,bool\,\rightarrow term\,bool$ 
  \item[] forall : $\Pi\alpha:type\rightarrow term(arr\,\alpha\,bool)\rightarrow term\,bool$

\end{itemize}



 

To translate the additional inference rules, four constants $\text{MP}$, $\text{DISCH}$, $\text{GEN}$ and $\text{SPEC}$ were introduced as below: 

\begin{itemize}
\item [] $\text{MP}$: $\Pi p:term\,\text{bool}.\Pi q:term\,\textit{bool.proof}(imp\,p\,q)\rightarrow \textit{proof}\,p\rightarrow \text{proof}\,q$
\item[] $\text{DISCH}$: $\Pi p:term\,\textit{bool}.\Pi q:term\,\textit{bool.proof}\,p\rightarrow proof\,q\rightarrow \textit{proof}(imp\,p\,q)$

\item[] $\textit{GEN}$: $\Pi\alpha:type.\Pi p':(term\,\alpha\rightarrow term\,bool).\Pi x:term\,\alpha.\textit{proof}\,(p'\,x)\rightarrow \textit{proof}(\textit{forall}\,\lambda\,x.p'x)$

\item[] $\textit{SPEC}$: $\Pi\alpha:type.\Pi t:(term\,\alpha\rightarrow term\,bool).\Pi u:term\,\alpha.\textit{proof}(\textit{forall}\,\alpha\,t)\rightarrow \textit{proof}(t\,u)$
\end{itemize}





 The translation of corresponding inference rules were added to Holide:

\begin{itemize}
\item [] \translate{\Omp} = ${\text{MP}|A||B||\mathcal{D}_{1}||\mathcal{D}_{2}|}$, where $\mathcal{D}_{1}$ and $\mathcal{D}_{2}$ are the proofs of $ A \rightarrow B$ and $A$ respectively.

\item[] \translate{\Ogen} = ${\text{GEN}|A||c'||\mathcal{D}'|}$,
  where $\,c'=\lambda x:||A||.|c|$, $\mathcal{D}$ is a proof of $A[c/x]$ and $\mathcal{D}'=\lambda x:||A||.|\mathcal{D}|$
  
 \item[] \translate{\Odisch} = ${\text{DISCH}|A||B||\mathcal{D}'||\mathcal{D}|}$, where $\mathcal{D}'$ is a proof of $A$ and $\mathcal{D}$ is a proof of $B$ 
 \item[] \translate{\Ospec} = $\text{SPEC}\,|A|t'|u||\mathcal{D}|$, where $t'=\lambda x:||A||.|t|$ and $\mathcal{D}$ is a proof of $B$.
\end{itemize}




\subsection{Evaluation}

 A way to compare proof size is to consider the size of the article files. To reduce the effect of syntax formatting and white-space, all the article files and Dedukti files  from both systems are compressed by \textit{gzip}. The size of both article files and Dedukti files scale down considerably after compression. Here we take the (OpenTheory) HOL Light's standard theory library for evaluation. As shown in Table \ref{size3}, the average size of the article files of HOLALA is around 64.36\% that of OpenTheory. This leads to an improvement of 41.81\% in translation time. The size of Dedukti files were reduced to about 64.92\% with an acceleration of 38.04\% for proof checking. 

\begin{table*}[!ht]
\centering
\caption{Comparison of Translation and Proof Checking}
\label{size3}
\begin{tabular}{l|p{5cm}|p{4cm}|}
\cline{2-3}
    & {Size of Proof Files (KB)} & {Translation Time (s)}  \\ \hline
\multicolumn{1}{|l|}{HOL Light} & 5,376                     & 55.98  \\ \hline
\multicolumn{1}{|l|}{HOLALA}  & 3,460                      & 32.57  \\ \hline
\multicolumn{1}{|l|}{Comparison}  &     Reduced to 64.36\%                  & Improved by 41.81\%  \\ \hline

\end{tabular}
\newline \vspace*{1 cm}
\begin{tabular}{l|p{5cm}|p{4cm}|}
\cline{2-3}
    & {Size of Dedukti Files (KB)} & {Proof Checking Time (s)}  \\ \hline
\multicolumn{1}{|l|}{HOL Light} &16,092                    & 30.75   \\ \hline
\multicolumn{1}{|l|}{HOLALA}  & 10,448                   & 19.05   \\ \hline
\multicolumn{1}{|l|}{Comparison}  &    Reduced to 64.92\%                 & Improved by 38.04\%  \\ \hline

\end{tabular}

\end{table*}



\section{Conclusion and Discussion}

An optimal design of a HOL kernel comes in various point of views: size and complexity, reasoning speed and memory efficiency, consideration of proof checking, etc. This paper presented HOLALA, a variant of HOL Light with an extended kernel by introducing implication and universal quantification. We provided an analysis on the reduction of the size of proofs and the reduction in time for proof checking. The size of proofs of HOLALA reduced to 64.36\% on average, leading to an improvement of a speed-up of 38.04\% for proof checking. It also worth noting that ITPs are usually developed without much concern about  the size of proofs and the complexity of proof checking. This paper attempted to bring theorem proving and proof checking closer with an emphasis on the efficiency of proof checking.  While OpenTheory grounds proofs to a minimal representation using a variant of HOL Light, this work shows the potential to ground proofs to a more efficient representation corresponding to a bigger (or the maximal) kernel instead. This work could be further completed by introducing conjunction and disjunction, truth and false, existential quantifier and more to the kernel. Another possible future work is to import proofs to (a variant of) HOL4 and export proofs out for further efficiency testing. Following this line, some further comparative experiments may be conducted between different extended kernels and the best efficiency payoff compared to its size. 

\section{Acknowledgement}
The author was supported by the MPRI-INRIA scholarship and greatly appreciated the supervision and kind help from Prof. Gilles Dowek, Dr. Ali Assaf, Dr. Joe Hurd, and Mr. Fr\'ed\'eric Gilbert on the understanding and implementation of HOLALA and Holide.

\bibliographystyle{unsrt}
\bibliography{main}

\end{document}

%% file: macpack.tex
\usepackage{graphicx}
\usepackage{etex}

\usepackage{listings}
\usepackage{standalone}

\usepackage{bmpsize}

\usepackage[table,xcdraw]{xcolor}
\usepackage{tikz}

\usepackage{booktabs}
\usepackage{multirow}

\usepackage{float}
\restylefloat{table}
\usepackage{amssymb}
\usepackage{tikz}

\usepackage{pbox}
\usepackage{amssymb}
\usepackage[document]{ragged2e}

\usepackage{amsmath,amsfonts,amssymb,amsthm,epsfig,epstopdf,url,array}

\usepackage{makeidx}
\usepackage{nomencl}

\usetikzlibrary{shapes.geometric, arrows}

\usepackage[T1]{fontenc}
\usepackage[latin9]{inputenc}
\usepackage{amsmath}
\usepackage{xargs}[2008/03/08]

\usepackage[english]{babel}

\usepackage{longtable}

\usepackage{afterpage}

\usepackage{pdfpages}
\usepackage{lipsum}
\usepackage{hyperref}
\usepackage{caption}
\usepackage{latexsym}
\usepackage{amssymb}

\usepackage{mathrsfs}

\usepackage{xspace}
\usepackage{graphics}
\usepackage{tikz}
\usepackage{mathtools}
\usepackage{bussproofs}

\EnableBpAbbreviations
\newcommand{\UICm}[1]{\UIC{$#1$}}
\newcommand{\AXCm}[1]{\AXC{$#1$}}
\newcommand{\BICm}[1]{\BIC{$#1$}}

\newcommand{\RLm}[1]{\RL{$#1$}}

\def\land{\wedge}
\def\lor{\vee}
\def\limp{\Rightarrow}
\def\fa{\forall}

\usepackage{setspace}

\usepackage[T1]{fontenc}
\usepackage{textcomp}
\usepackage{listings}

\usepackage{amssymb}

    \newcommand{\HOLASSUME}{
    \AXCm{~} \RLm{ASSUME}
    \UICm{\left\{{A}\right\} \vdash A}
    \DP}

\newcommand{\HOLMPl}{
    \AXCm{\Gamma \vdash A \limp B} \AXCm{\Delta \vdash A}
    \RLm{MP} 
    \BICm{\Gamma \cup \Delta \vdash B} \DP}
    
\newcommand{\HOLDISCHl}{
    \AXCm{\Gamma \vdash B}
    \RLm{DISCH}
    \UICm{\Gamma \setminus \left\{ {A}\right\} \vdash A \Rightarrow B} \DP}

\newcommand{\HOLGENl}{
    \AXCm{\Gamma \vdash A[c/x]} \RLm{GEN$ if x is not free in $ \Gamma}
    \UICm{\Gamma \vdash \fa{x} A} \DP
    }

\newcommand{\HOLSPECl}{    
    \AXCm{\Gamma \vdash \fa{x}A} \RLm{SPEC} 
    \UICm{\Gamma \vdash A[t/x]  } \DP
    }

\newcommand{\HOLREFL}{
    \AXCm{~} \RLm{REFL}
    \UICm{ \vdash A = A}
\DP}

\newcommand{\HOLTRANS}{
 \AXCm{\Gamma \vdash A = B} \AXCm{\Delta \vdash B = C} \RLm{TRANS}
    \BICm{\Gamma \cup \Delta \vdash A = C}
\DP
}

\newcommand{\HOLMKCOMB}{
\AXCm{\Gamma \vdash A = B} \AXCm{\Delta \vdash C = D} \RLm{MK\_COMB}
\BICm{\Gamma \cup \Delta \vdash A (C) = B (D)}

\DP}

\newcommand{\HOLABS}{
    \AXCm{\Gamma \vdash A = B}
    \RLm{ABS}
    \UICm{\Gamma \vdash \lambda x. A = \lambda x. B}
    \DP
}

\newcommand{\HOLBETA}{
    \AXCm{~}
    \RLm{BETA}
    \UICm{(\lambda x. A) x = A}
\DP}

\newcommand{\HOLEQMP}{
    \AXCm{\Gamma \vdash A = B }
    \AXCm{\Delta \vdash A}
    \RLm{EQ\_MP}
    \BICm{\Gamma \cup \Delta \vdash B}
\DP}

\newcommand{\HOLDEDUCTAS}{
    \AXCm{\Gamma \vdash A}
    \AXCm{\Delta \vdash B}
    \RLm{DEDUCT\_ANTISYM\_RULE}
    \BICm{(\Gamma \setminus \left\{ {B}\right\}) \cup 
    \Delta \setminus \left\{ {A}\right\}) \vdash A = B}
\DP}

\newcommand{\HOLINST}{
    \AXCm{\Gamma [x_1, \dots, x_n] \vdash A[x_1, \dots, x_n]}
    \RLm{INST}
    \UICm{\Gamma [t_1, \dots, t_n] \vdash A[t_1, \dots, t_n]}
\DP}

\newcommand{\HOLINSTTYPE}{
    \AXCm{\Gamma [\alpha_1, \dots, \alpha_n] \vdash A[\alpha_1, \dots, \alpha_n]}
    \RLm{INST\_TYPE}
    \UICm{\Gamma [\gamma_1, \dots, \gamma_n] \vdash A[\gamma_1, \dots, \gamma_n]}
\DP}

\newcommand{\Omp}{
\AXCm{\Gamma \vdash A \limp B} \AXCm{\Delta \vdash A}
\RLm{MP}
\BICm{\Gamma \cup \Delta \vdash B} \DP
}

\newcommand{\Odisch}{
\AXCm{\Gamma \vdash B}
    \RLm{DISCH} 
    \UICm{\Gamma \setminus \left\{ {A}\right\} \vdash A \Rightarrow B} \DP
}

\newcommand{\Ogen}{
   \AXCm{\Gamma \vdash A[c/x]} \RLm{GEN $\, if x is not free in $\Gamma}  
    \UICm{\Gamma \vdash \fa{x} A} \DP
}

\newcommand{\Ospec}{
  \AXCm{\Gamma \vdash \fa{x}A} \RLm{SPEC}
    \UICm{\Gamma \vdash A[t/x]  } \DP
}

\newcommand{\vbar}{$\biggr\rvert$}

\newcommandx\prove[3][usedefault, addprefix=\global, 1=]{\AXCm{#2}\RL{#1}\UICm{#3}\DP}

\newcommandx\translate[1][usedefault, addprefix=\global]
{{\vbar}{#1}{\vbar}}

\usepackage{multicol}

\usepackage{lscape}

\newcommand{\examplehol}{
\begin{prooftree}

\AXCm{}
\RLm{ASSUME}
\UICm{p \land  (p\Rightarrow q) \vdash p \land (p\Rightarrow q)}
\RLm{CONJUNCT2}
\UICm{p \land  (p\Rightarrow q) \vdash p\Rightarrow q}

\AXCm{}
\RLm{ASSUME}
\UICm{p \land  (p\Rightarrow q) \vdash p \land (p\Rightarrow q)}
\RLm{CONJUNCT1}
\UICm{p \land  (p\Rightarrow q) \vdash p}

\RLm{MP}
\BICm{p \land  (p\Rightarrow q) \vdash q} 
\RLm{DISCH}
\UICm{\vdash (p \land (p \Rightarrow q)) \Rightarrow q}

\DP
}